\documentclass[twoside,twocolumn,fleqn,10pt]{wlscirep}
\usepackage[utf8]{inputenc}
\usepackage[T1]{fontenc}
\usepackage{siunitx}
\usepackage{multirow}

\title{Portable ground stations for space-to-ground quantum key distribution}

\author[1,2,3,*]{Ji-Gang Ren}
\author[1,2,3,6*]{Maimaiti Abulizi}
\author[4,*]{Hai-Lin Yong}
\author[1,2,3]{Juan Yin}
\author[1,2,3]{Xue-Jiao Li}
\author[1,2,3]{Yuan Jiang}
\author[1,2,3]{Wei-Yang Wang}
\author[1,2,3]{Hua-Jian Xue}
\author[1,2,3]{Yu-He Chen}
\author[4]{Biao Jin}
\author[1,2,3]{Ya-Yun Yin}
\author[1,2,3]{Zhou-Yu Tu}
\author[4]{Xiao-Juan Zhu}
\author[4]{Shuang-Qiang Zhao}
\author[1,2,3]{Feng-Zhi Li}
\author[1,2,3]{Sheng-Kai Liao}
\author[1,2,3]{Wen-Qi Cai}
\author[1,2,3]{Wei-Yue Liu}
\author[1,2,3]{Yuan Cao}
\author[5]{Fei Zhou}
\author[1,2,3]{Li Li}
\author[1,2,3]{Nai-Le Liu}
\author[1,2,3]{Qiang Zhang}
\author[1,2,3]{Yu-Ao Chen}
\author[1,2,3,7]{Cheng-Zhi Peng}
\author[1,2,3,8]{Jian-Wei Pan}

\affil[1]{Hefei National Laboratory for Physical Sciences at the Microscale and Department of Modern Physics, University of Science and Technology of China, Hefei 230026, China}
\affil[2]{Chinese Academy of Sciences (CAS) Center for Excellence and Synergetic Innovation Center in Quantum Information and Quantum Physics, University of Science and Technology of China, Shanghai 201315, China}
\affil[3]{Shanghai Research Center for Quantum Sciences, Shanghai 201315, China}
\affil[4]{QuantumCTek Co., Ltd. , Hefei 230088, China}
\affil[5]{Jinan Institute of Quantum Technology, Jinan 370102, China}
\affil[6]{School of Physics and Electronic Engineering, Xinjiang Normal University, Urumqi 830054, China}

\affil[7]{Corresponding author: pcz@ustc.edu.cn}
\affil[8]{Corresponding author: pan@ustc.edu.cn}

\affil[*]{These authors contributed equally to this work}

\begin{abstract}
Quantum key distribution (QKD) uses the fundamental principles of quantum mechanics to share unconditionally secure keys between distant users.
Previous works based on the quantum science satellite ``Micius'' have initially demonstrated the feasibility of a global QKD network.
However, the practical applications of space-based QKD still face many technical problems, such as the huge size and weight of ground stations required to receive quantum signals.
Here, we report space-to-ground QKD demonstrations based on portable  receiving ground stations.
The weight of the portable ground station is less than 100\,kg, the space required is less than 1\,m$^{3}$ and the installation time requires no more than 12 hours, all of the weight, required space and deployment time are about two orders of magnitude lower than those for the previous systems. Moreover, the equipment is easy to handle and can be placed on the roof of buildings in a metropolis. 
Secure keys have been successfully generated from the ``Micius'' satellite to these portable ground stations at six different places in China, and an average final secure key length is around 50 kb can be obtained during one passage. 
Our results pave the way for, and greatly accelerate the practical application of, space-based QKD. 
\end{abstract}

\begin{document}
\flushbottom
\maketitle
%
%


\section{Introduction}
Information security is a crucial component of communication networks in the information era.
The security of widely used classical cryptosystems, such as the RSA public-key cryptography \cite{rivest1978}, are based on high complexity of mathematical problems.
A quantum computer running Shor's factoring algorithm \cite{ShorAlgorithm_94,shor1999} could potentially crack these encryption systems.
Based on the principles of quantum mechanics, QKD provides a solution to extending the length of a secure symmetric key.

\begin{figure*}[htbp]
\centering
\includegraphics[width=0.85\textwidth]{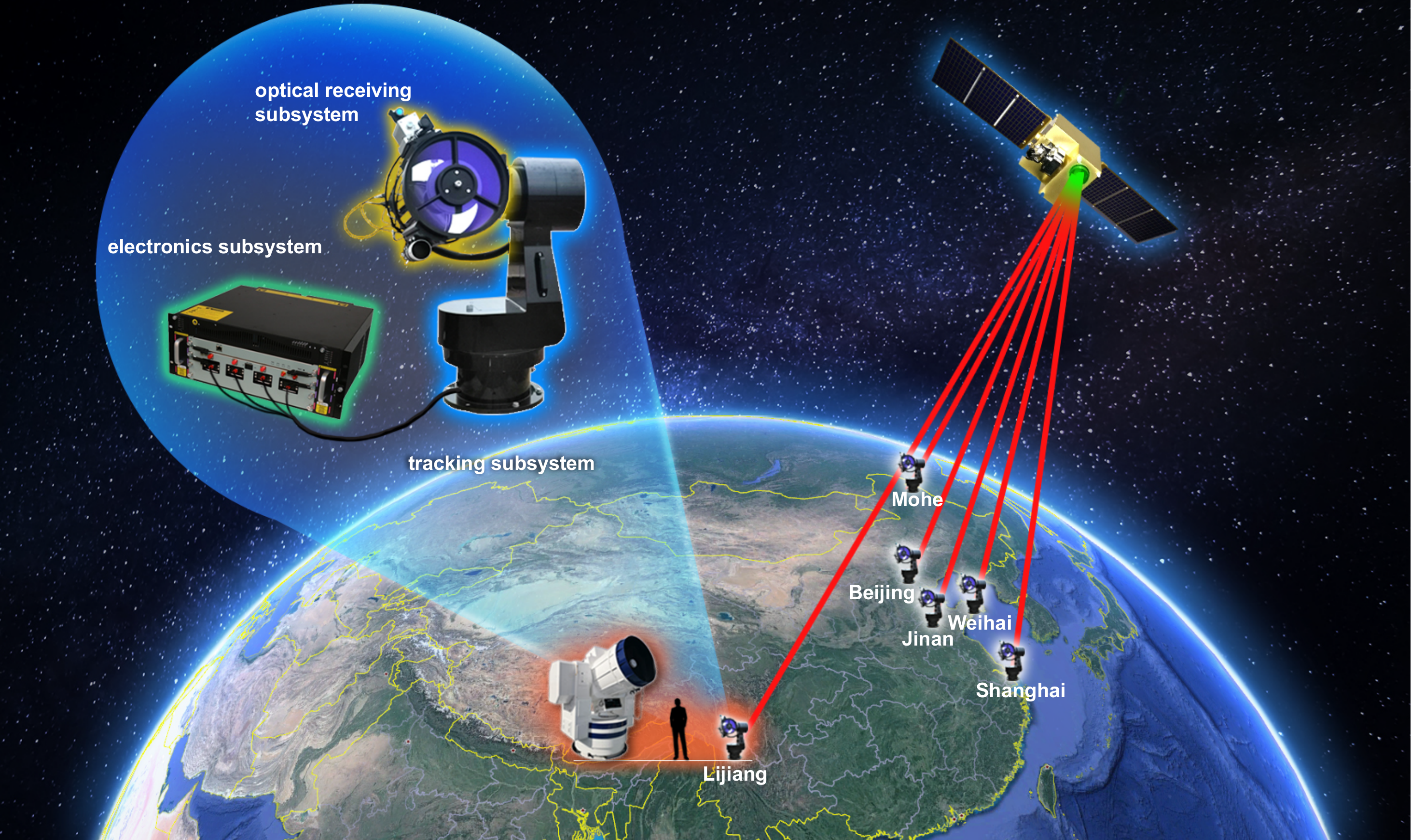}
\caption{Overview of space-to-ground quantum key distribution based on portable ground stations. The portable ground stations are deployed at six places, including Shanghai, Beijing, Jinan, Weihai, Mohe and Lijiang. The portable ground station consists of an optical receiving subsystem, a tracking subsystem and an electronics subsystem. The received photons coupled into the optical receiving subsystem and are transferred to the detectors in the electronics subsystem via multi-mode fibers. The image of the Earth is from Google.}
\label{fig:scheme}
\end{figure*}

The first QKD protocol (the BB84 protocol) was proposed in a seminal paper by Bennett and Brassard in 1984 \cite{Bennett:BB84:1984}.
From then on, many demonstrations have been carried out via fiber and free-space channels, and the secure keys have been distributed over longer and longer distances \cite{Muller:23km:1996,Hughes2002:10km,GYS:122km:2004,Zeilinger:Decoy:2007,Wang:Direct:2013,Boaron:421km:PRL:2018}.
Currently, the maximum ground-based distribution range achieved through a fiber is \SI{509}{\kilo\meter} \cite{JPChen:509km:PRL:2020}.
Satellite-to-ground QKD and a quantum-secured video conference between Beijing and Vienna were carried out with the ``Micius'' satellite in 2017 \cite{Liao:SatQKD:2017,Liao:2018:intercontinentalQKD,Yuao:SatBackbone:nature:2021}.
These groundbreaking results have spurred an international enthusiasm for applications of space-based quantum communications \cite{takenaka2017satellite,pugh2017airborne,schimmel2018free,Wang:OE:FSCVQKD:2020,bedington2017progress,Pirandola_2020,anisimova2017mitigating,chandrasekara2017tracking}.

Currently, many research groups are working on space-based QKD missions to establish the global quantum communication networks\cite{2018:Italy:SatelliteGNSS,2020:UK:Schedule}.
Many space missions are in progress, including QEYSSAT \cite{Satellite:Canada:NJP2013}, QUARTZ \cite{2018:QUARTZ:SES}, QUBE \cite{2018:Haber:DLR:Qube}, SpooQySats \cite{2018:Singapore:SpooQySats}, NanoBob \cite{2018:EU:nanobob}, Q\textsuperscript{3}Sat \cite{2018:UK:Q3}, and 
CAPSat \cite{2016:CAPSat:IllinoisNASA}.
However, to make the space-based QKD techniques practical, many unsolved problems remain to be solved in the application scenario.
The most significant problem is to develop portable ground stations.
The four optical receiving optical ground stations currently used for ``Micius'' are mainly based on giant telescopes.
All these telescopes \cite{Liao:SatQKD:2017,2018:He:AO:telescope} have apertures greater than \SI[number-unit-product=-]{1}{meter} aperture, and they each weigh more than \SI{10}{tons}.
Such giant telescopes also need to be placed in professional astronomical domes, which are always located in an observatory station on the top of a hill remote from urban areas.
System installation, commissioning, and calibration also take months and even years\cite{TMT:timeline,1m_telescope:timeline}, and these difficulties greatly limit the practicality of space-based QKD.

Application-oriented next-generation quantum satellites are expected to be as small as a micro- or nano-satellites, which have lower development and launch costs \cite{2018:Singapore:SpooQySats,2018:EU:nanobob,2018:UK:Q3,2020:UK:Schedule}. 
Similarly, in order to achieve large-scale deployment, the optical ground stations must be user-oriented and miniaturized.
Based on long-term exchanges of views with potential users, we find that a practical ground station must meet the following conditions: Firstly, it should be light and small enough to be transported by an elevator without requiring cranes or forklifts. Second, it should be possible to place it on a normal flat roof of a common building without needing special requirements.
Finally, the installation, commissioning, and operation process should be simple and clear, so that users without professional backgrounds will be able to operate the equipment.

Here, we describe the development of a portable quantum communication ground station that meets these specific requirements of users. It can receive downlink signal photons efficiently from the ``Micius'' satellite, so we can share quantum keys between satellite and ground. %
Because it uses a telescope with a \SI[number-unit-product=-]{280}{\milli\meter} aperture instead of a much larger one, the portable ground station weighs only \SI{100}{\kilo\gram}, with power consumption less than 300 Watts and size less than 1 m$^{3}$. This station can be moved quickly to any location the user wants.
However, due to the smaller telescope aperture, channel attenuation is increased, while the background noise does not change much. This system configuration leads to a lower signal-to-noise ratio, a lower raw key rate, and higher quantum bit error rate(QBER), and these variations limit the performance of the portable ground station.
According to previous QKD experiments with ``Micius'' \cite{Liao:SatQKD:2017,Liao:2018:intercontinentalQKD}, the sifted key rate at a distance of \SI{1,000}{\kilo\meter} can only reach the order of \SI{100}{bps} with a portable ground station, making it less likely to obtain secure keys due to finite key size effects. To solve this problem, we have improved the receiving efficiency and polarization fidelity through an integrated optical design. The system efficiency and polarization fidelity can reach \SI{40}{\percent} 
and \SI{99.5}{\percent}, respectively, with such a receiving system. The background noise is reduced by using a narrower filter, and we have been able to realize QKD even in the strong background light environments in a cities.
In our system, the receiving field of view for quantum signals is designed to be \SI{100}{\micro\radian}, which reduces the requirements for pointing and tracking accuracy.
Therefore, one can remove the fine-tracking system, enabling the ground station to maintain a higher optical efficiency. This greatly reduces the complexity of the system, which consequently reduces the time required for installation and testing of the ground station.
From joint satellite-ground measurement, we obtained a sifted key rate around \SI{2}{kbps} for 500 - 1000 km satellite link with such a smaller aperture telescope.
The entire process from the beginning of unpacking, through installation, to tracking the satellite does not exceed \SI{12}{hours}.
We believe that the development of \SI{880}\times\SI{480}\times\SI{480}{\milli\meter}, and successful experiments with, these portable quantum communication ground stations is an important aspect of the practical application of space quantum communication.

\section{System architecture}
\label{sec:design}
The feasibility of uplink and downlink quantum communications have been discussed and compared in previous works \cite{bonato:feasibility:2009,Satellite:Canada:NJP2013}.
The uplink is more susceptible to the atmospheric environment \cite{Ren:SatTele:2017,Han:2018:point-ahead}. The lifetime of single-photon detectors with low dark counts is also a challenging technical difficulty, which requires special protection in space due to the space-radiation environment \cite{MYang:detector:OE:2019}.
Therefore, an optical downlink with smaller attenuation is a reliable choice for practical use in space quantum communication. 
In this downlink scenario, the quantum satellite acts as an optical transmitter, while the ground stations are the receivers \cite{Liao:SatQKD:2017,Liao:Tianggong:CPL:2017,Yuao:SatBackbone:nature:2021}.

At the satellite part, the photons are encoded by four different linear polarization (H/V/+/-) and three intensity levels due to decoy-state protocol. The repeat frequency of the source is 100 MHz and the average photon number in the output of the telescope is: $\mu_s$ = 0.8 (signal state), $\mu_d$ = 0.1 (decoy state) and $\mu_0$ = 0 (vacuum state), with probabilities of 50\%, 25\% and 25\%, respectively. 

A portable ground station, as shown in Fig.~\ref{fig:scheme}, consists of optical receiving, tracking, and electronics subsystems.
The optical receiving subsystem is used to collect photons from the satellite and couple them into optical fibers, while the tracking subsystem controls the optical receiving subsystem to point and track the satellite, which maintains the satellite-to-ground optical link.
The electronics subsystem converts the arrived single-photon signals to time-tagged data, and subsequently performs the data post-processing according to the decoy-state QKD protocol.

\subsection{Optical receiving subsystem}
\begin{figure}[tb]
\centering
\includegraphics[width=1.0\linewidth]{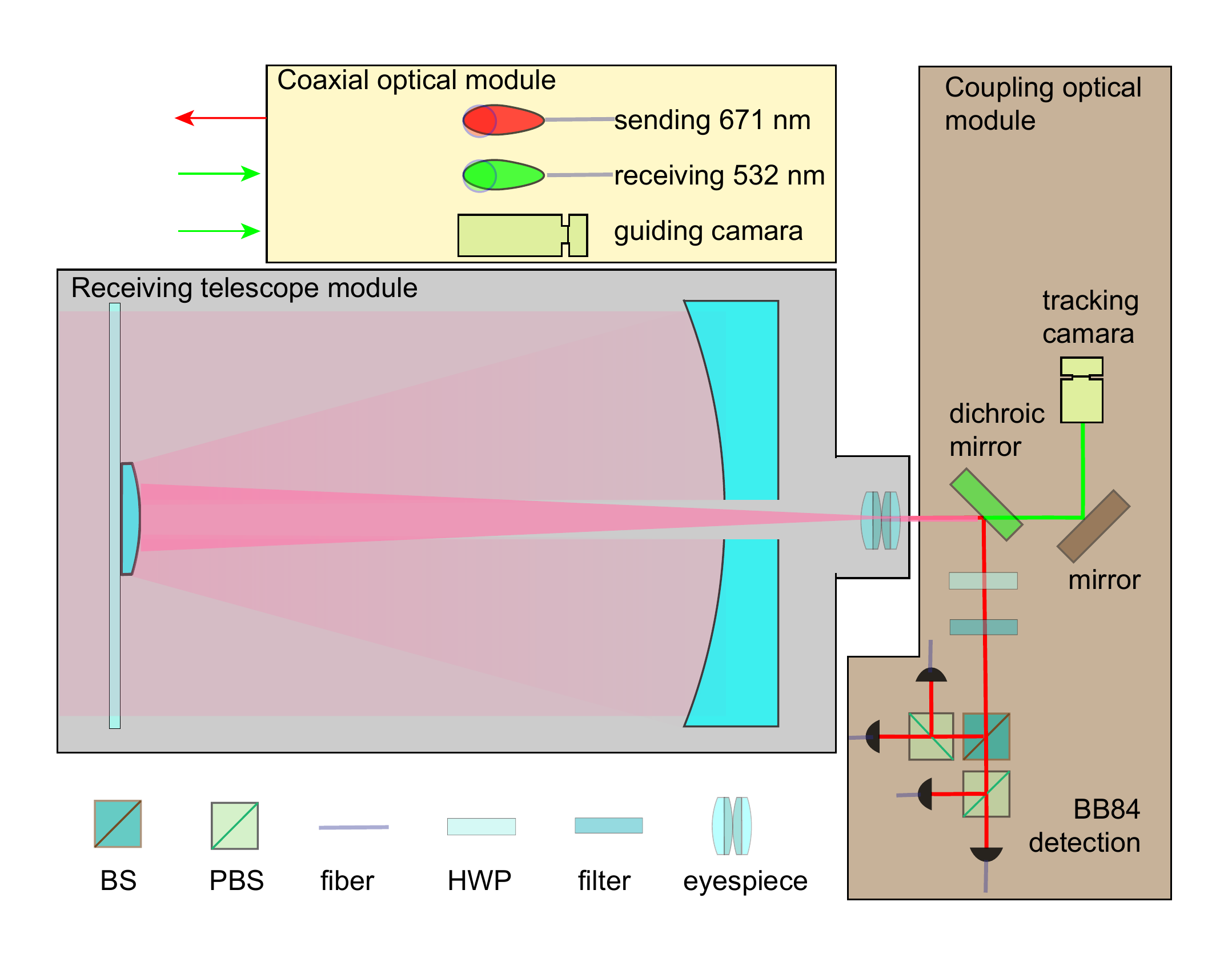}
\caption{A schematic diagram of the optical receiving subsystem. This optical receiving subsystem is composed of three modules, including a receiving telescope, a coupling optical module and a coaxial optical module. The telescope is a Schmidt-Cassegrain type one with \SI{280}{\milli\meter} in diameter. The coupling optical module is fixed onto the back of the telescope. The narrow-band interference filter is used to reduce the noisy photons. The coaxial optical module is installed on one of the dovetail plates of the telescope. As output, the received signal photons and synchronizing light signals are all transmitted to the electronics subsystem via multi-mode fibers. BS: beamsplitter, PBS: polarization beamsplitter, HWP: half-wave plate.}
\label{fig:opticalschematic}
\end{figure}

The most important part of the ground station is the optical receiving subsystem. It is composed of a receiving telescope, a coupling optical module, and an auxiliary coaxial optical module, as shown in Fig.~\ref{fig:opticalschematic}.
The receiving telescope has a Schmidt-Cassegrain type structure; it is a \SI{280}{\milli\meter} in diameter and \SI{531}{\milli\meter} length.
Photons from the satellite that impinging on the ground station are collected by the primary and secondary mirrors and converge at the focal point. 
A high-order, aspheric Schmidt plate at the entrance of the telescope is used to correct the spherical aberration. 
After the focal point, photons are collimated with a Pl\"ossl eyepiece and guided to the BB84 measurement module.
Better than the usual amateur telescopes, the reflecting and transmitting surfaces of the telescope are coated with dielectric films to ensure high optical efficiency greater than \SI[number-unit-product={}]{90}{\percent} in the \SI{850}{\nano\meter} band.
For the benefit of system pointing calibration and self-testing, the total efficiency of the telescope for the visible light (\SIrange[range-phrase=--,range-units=single]{390}{780}{\nano\meter}) is designed to exceed \SI[number-unit-product={}]{80}{\percent}.

\begin{figure}[tb]
\centering
\includegraphics[width=1.0\linewidth]{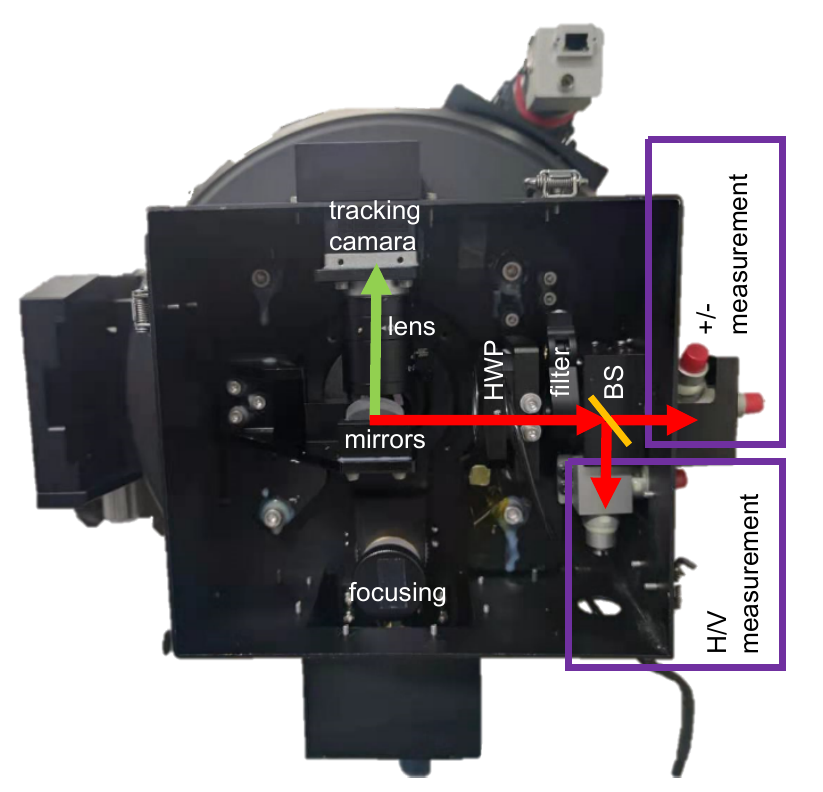}
\caption{A photo of the coupling optical module of the optical receiving subsystem. After the receiving telescope, the photons are transferred into the coupling optical module. First, the beacon light is reflected to the tracking camera as the tracking input source (green line), while the 850 nm photons are reflected into the BB84 measurement components (red line). The +/- measurement part is rotated by 45 degrees around the optical axis instead of using the extra HWP. A moter-driven HWP is used to compensate for the angular shift between the satellite and the frame of the ground station. To ensure the temperature adaptability, the focusing structure is used only when the temperature changes greatly. A flipping filter is used to lower the background photon noise.}
\label{fig:opticalmodule}
\end{figure}

As shown in Fig.~\ref{fig:opticalmodule}, the entire coupling and measuring optical module is directly fixed onto the back of the telescope. 
All the optical components of this module have dimensions less than \SI{25}{\milli\meter} in dimensions, which provides a sufficient clear aperture and field of view.
The light and stable base board and opto-mechanical mounts are made of aluminum, with a compact layout and lightweight design.
Most of the devices are actually located within the envelope of the telescope diameter.

The photons from the eyepiece are collimated into a beam with a diameter of \SI{\sim4.5}{\milli\meter}.
Then, as shown in Fig.~\ref{fig:opticalmodule}, the beam is divided into two paths by a \ang{45} polarization-maintained dichroic mirror.
A \SI[number-unit-product=-]{532}{\nano\meter} beacon-laser beam is transmitted to the camera in the coupling optical module for more accurate angular offset measurement, while the \SI[number-unit-product=-]{850}{\nano\meter} photons are reflected.
Those photons subsequently pass through a compensating HWP and a narrowband filter, enter the BB84 measurement module, and then are coupled into multi-mode optical fibers that are connected to four single-photon detectors.
To reduce the difficulty of integrating the subsystem and increase the deviation tolerance for practical applications, the field of view and effective aperture of the BB84 module are designed to be \SI{6.5}{\milli\radian} and \SI{6}{\milli\meter}, respectively, which have sufficient margins compared to the input beam.
The efficiency of the dichroic mirror (DM) and HWP used in the module are greater than \SI[number-unit-product={}]{99}{\percent}, and the transmittance of the narrow band-pass filter, which has a \SI{1}{\nano\meter} bandwidth centered at \SI{850}{\nano\meter}, is greater than \SI[number-unit-product={}]{92}{\percent}.
Finally, the coupling efficiency to the multi-mode fibers, (which have core diameters of \SI{105}{\micro\meter} and numerical aperture of \num{0.22}) is greater than \SI[number-unit-product={}]{80}{\percent}.
The measured total efficiency from the entrance to the coupling fibers in the optical receiving subsystem is higher than 60\% in the laboratory.

In the optical subsystem, the direction of the polarization does not change with the orientation of the telescope. 
However, we still need to configure a rotatable HWP in the quantum receiving module to compensate for the change in polarization-basis direction caused by satellite motion.
The rotation trajectory of the HWP can be calculated from the satellite orbit data.
During satellite tracking, the HWP rotates in an open-loop style with a maximum drift \ang{<0.3}.
The rotating device can be removed if the H-polarization state emitted by a next-generation satellite is always aligned with the horizontal plane.
The total polarization error is measured to be \SI[number-unit-product={}]{<0.2}{\percent} for the two pairs of orthogonal polarization states of the BB84 protocol.

In an auxiliary optical module, three components are installed coaxially on the dovetail plates of the telescope.
To acquire the beacon laser, a guiding camera with a field-of-view larger than \ang{1} is installed on one of the dovetail plates. 
A second dovetail board contains the beacon laser, with a wavelength of \SI{671}{\nano\meter}, an output power \SI{1}{\watt}, and a divergence angle of  \SIrange[range-phrase=--,range-units=single]{2}{4}{\milli\radian}, which aids in satellite tracking at the ground station.
The beacon light from the satellite is a pulsed green laser with 10 kHz frequency, whose time tags for synchronization are logged in the electronic device in the satellite.
A small fiber collimator is installed after a 532 nm band-pass filter on the third dovetail plate to couple a small portion of the beacon laser beam from the satellite as a synchronizing signal.
A single-photon detector connected to a multi-mode fiber is used to detect the synchronizing signal. As a result, the time synchronization accuracy is less than 1 ns.
The non-parallelism between the auxiliary optical module and the telescope is smaller than \SI{0.5}{\milli\radian}, which is not too difficult to deal with as the stress and temperature change.

\subsection{Tracking subsystem}
The tracking subsystem in the ground station is composed of an integrated control computer with software, and a tailored two-dimensional direct-drive mount with a servo controller.
The direct-drive mount uses an L-shaped frame structure with elevation axis (\ang{-3},\ang{93}) and an azimuth axis (\ang{-270},\ang{270}), see the appearance of the mount in Fig.~\ref{fig:scheme}. The optical receiving subsystem is mounted on the elevation axis.
Compared with a traditional U-shaped frame, the L-shaped frame has the advantage of small size, with dimensions of $\Phi 250 mm\times H 850 mm$, and it weighs about \SI{70}{\kilo\gram}.
Taking into account the \SI[number-unit-product=-]{100}{\micro\radian} field of view, a closed-loop tracking error of \SIrange[range-phrase=--,range-units=single]{10}{15}{\micro\radian} is acceptable and is already easily achieved.
Unlike the giant telescopes used previously, no fast-steering mirrors are needed for high-precision and high-frequency tracking.

Each axis is driven by a direct current torque motor and is supported by a pair of high-precision bearings.
As an angle measuring device, we use high-precision circular gratings with \SI[number-unit-product=-]{26}{bit} absolute encoders, which enables us to achieve a resolution of up to \ang{;;0.02}.
The servo tracking system mainly drives the telescope to complete the target-tracking task based on the results from processing the sensor data in the previous stage.
To track a space target, the torque motors are required to rotate in a very precise way in both the azimuth and elevation angles. The control mode of the servo-tracking system is a double closed-loop control structure.
The inner ring is a speed loop for correcting the angular velocity, and the outer ring is a position loop for correcting the angular direction of the mount.

Connected through network cables, the control computer can control various parts of the ground station, including the cameras, the dual-axis mount, the polarization-compensating HWP, and a GPS timer.
Using the camera-control unit in the control software, the operators can adjust the exposure time and gain of the guiding and tracking cameras in the coupling optical module, and can choose automatic or manual control, according to the working modes and target characteristics. After acquiring an image, the data-processing unit performs a series of numerical image-processing steps, including image filtering, target contrast enhancement, and background suppression. This unit finally returns the offset angles of the target and sends them to the servo unit. The servo unit fine-tunes the direction of the mount to realize the closed-loop position control.

After loading the forecast data of the satellite orbit into the software, the telescope mounted on the mount points to the initial position of the satellite (usually at an elevation of \ang{5}).
When the satellite passes by on its track, the ground station drives its beacon laser to cover the satellite \cite{Liao:SatQKD:2017}.
Once the guiding camera acquires the satellite's laser beacon, the control computer couples the laser beacon into the optical coupling module and adjusts the direction of the telescope to point to the satellite accurately, using the offset angle obtained from the tracking camera.
Once a stable optical link for quantum communication has been established, the photons are coupled into the fibers in the optical receiving subsystem and are then sent to the electronics subsystem.

\subsection{Electronics subsystem}

The electronics subsystem of the ground station system is contained in a 19-inch standard case with a 4U chassis. This subsystem contains five single-photon avalanche diodes (SPADs) and an FPGA-based multi-channel TDC, which are placed separately next to the mount. Four multimode fibers carrying quantum optical signals and one multi-mode fiber carrying the synchronizing optical signals are connected to the detectors in the electronics subsystem. The cable-wrap problem caused by rotation of the mount is solved by the design of the tracking subsystem. The fibers connecting the optical subsystem and the detection module are fixed through this cable channel.

The photons from the fibers are detected by the SPADs and converted into electrical signal pulses.
Pulse per second (PPS) signals are also recorded in another channel of the TDC for matching the time starting points. The arrival times of the electrical signals can be measured accurately by the TDC —to an accuracy of about \SI{50}{\pico\second}—and finally all measurement events are recorded in the recording computer. With the recorded data and the information obtained at radio stations via classical communication at radio frequencies from the satellite, secure keys can be generated through the software according to the decoy-state QKD protocols.
Automatic post-processing of the data can be integrated into and performed in this programmable electronics subsystem. In future commercial applications, optical communication can be used as a classical communication channel, which further accelerates the efficiency of key generation.

\section{Demonstrations in the Field}

In order to verify the performance, stability, and convenience of the portable ground station, we chose six locations to carry out the experiments in QKD based on ``Micius'', as shown in Fig \ref{fig:scheme}.
The first two places were a close suburb of Shanghai and a central area of Beijing, and they are two large metropolitan areas with strong background light. The third and fourth places were in two typical cities in eastern China, the inland city Jinan and a beachside area in the coastal city Weihai that has relatively high water vapor.
The fifth location was Mohe, Heilongjiang Province, the northernmost village in China, which is at a high latitude, so the satellite is likely to be illuminated by the sun.
The sixth place was a high-altitude astronomical observatory with very good air quality in Lijiang, Yunnan Province. The selection of these experimental sites covers different environmental characteristics and application scenarios, which can fully verify the performance of the portable ground station. 

A typical deployment process is as follows: first, the equipment is transported to the place required by the user.
When the installation has been completed, we carry out parameter calibrations, such as zero corrections, pointing-error corrections, and channel-efficiency evaluations.
Finally, we perform the QKD experiment with ``Micius'', sharing the quantum key between the ground and the satellite.

\subsection{Rapid assembly and deployment}

We reduced the total weight of the ground-station equipment to less than 100 kg, and the ground station can be packaged into three parts when transportation is needed. The first part is the modified optical telescope, the second part is the mount, and the third part is the electronics and equipment cabinet. When the equipment arrives at the specified location, we first install the telescope onto the mount, test the balance, and adjust the base of the mount near horizontal and make its zero direction about due north (for example, using mobile phones to set the azimuth accuracy to within 10 degrees). We then connect the telescope and mount and connect the cables to the equipment cabinet. Next, we set the parameters and related software and boot the system to confirm that the equipment is basically functioning normally. This process takes about two or three hours of work by three people, and it can be done during the day or night. At night, we calibrate the model and test the system efficiency. After this, the equipment is capable of conducting experiments with ``Micius''.

The first thing to be calibrated is the direction offset angle of the mount, because whether the mount is pointing to a satellite or a star, the angle of the mounting frame itself needs to be calculated according to the position of the target in an Earth-centered inertial coordinate frame. However, in the process of ground-station equipment placement, only simple tools such as a compass are used for alignment. This ensures that when the mount points to a star, its light may not enter the field of view of the capture camera. Therefore, it is necessary to find any bright star at night and acquire it manually on both the capture camera and the tracking camera at the same time. Then the mount is controlled to move the center of the star-image spot to the calibration position of the tracking camera, which represents the position with the highest receiving efficiency for the quantum signal. At this time, the actual position of the mount can be recorded, and the zero deviation of the two frames can be corrected.

\begin{figure}[htbp]
\centering
\includegraphics[width=0.9\linewidth]{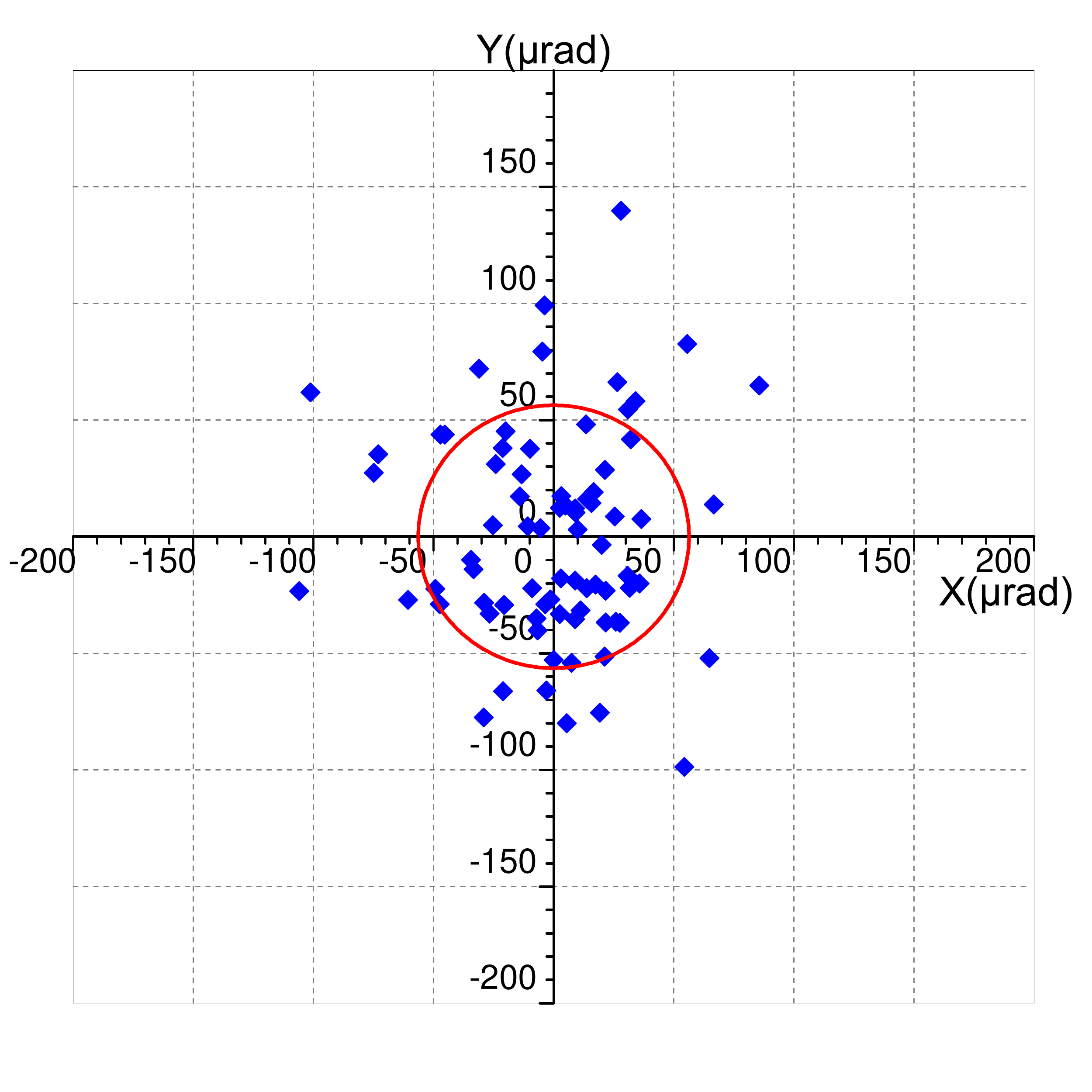}
\caption{Pointing errors to 78 stars after quick calibration with several stars on August, 2018. Each blue point shows the shift between the angular position of the star and the calibrated model. The RMS error of the pointing angle (red circle) to these stars is \SI{56.36}{\micro\radian}.}
\label{fig:point}
\end{figure}

After the direction offset angle calibration,  pointing the telescope to any star can bring the star image into the field of view of the tracking camera. However, due to the influence of the gravitational-deformation error of the telescope, the non-orthogonality of the altitude axis and the azimuth axis, the tilt of the azimuth axis, etc., when the telescope points to different stars or to different directions of the satellite, the pointing position may have a large deviation from the actual position. In this case, the beacon light installed on the telescope cannot cover the satellite, so we cannot achieve two-way, closed-loop tracking. In order to solve this problem, we use the correction method in the pointing model proposed by Wallace \cite{2002:Wallace:Pointing},  employing 15–20 stars spread over the whole sky to correct for the pointing error of the telescope itself and for atmospheric-transparency differences (e.g., varying cloudiness). After this correction, when the telescope points to different stars in the whole sky again, the correction accuracy can be obtained by measuring the relative deviation of the spot center.
This work needs to be carried out on a clear night, and the whole correction process needs about an hour. Fig.~\ref{fig:point} shows the results of a pointing-model experiment in Shanghai station on August, 2018. The pointing accuracy of different stations is related to the number of stars selected and the status of the equipment. Generally, if more than 20 stars are selected, the pointing errors do not exceed \SI{100}{\micro\radian}.

Obtaining a high SNR plays a crucial role in space quantum communication. The SNR is related to channel efficiency and noise counts. Channel efficiency includes both atmospheric attenuation and the efficiency of the receiving telescope. The noise counts includes dark counts of the detectors and background counts. So, it is necessary to determine quickly whether the channel efficiency and noise counts meet the experimental requirements before the satellite crosses the horizon. We can verify the efficiency of the ground-station system by directing the telescope to a star, and we can test the effect of noise in different directions by facing different positions in the sky.

The channel efficiency of the ground station is seriously limited by the weather conditions.
In an efficiency test based on stars, we assume that the stars will send the corresponding density of photon counts to the Earth according to the value of the I-band magnitude.
The total efficiency of the ground station includes the atmospheric loss caused by the angle of the star and weather factors.

In a rural environment without a moon and on a clear night, the total background-noise-induced single-photon counts for the entire system can be less than 100 cps. In the center of most cities and in a full-moon environment, however, the total noise counts maybe around 2000 cps when the elevation angle of the mount is less than \SI{15}{\degree}, although the noise counts decrease rapidly when the telescope points to elevations greater than  \SI{15}{\degree}. When the elevation is greater than \SI{30}{\degree}, the total noise count of a detector is usually close to that obtained in a rural environment.

\subsection{Satellite-to-ground QKD results}

After the pointing model, efficiency and noise level are confirmed to be qualified, the mission of a satellite-to-ground QKD experiment can be executed with this ground station.
For a specific day, due to the different weather conditions at the available ground stations, the satellite planning center provides precise satellite-orbit predictions before the middle of the day, and sends the instruction sequence to the satellite through the controlling center. In the afternoon, the trajectory file for the mount of the ground station is generated at the planning center and is then sent out to the corresponding ground station.

\begin{figure}[htbp]
\centering
\includegraphics[width=1.0\linewidth]{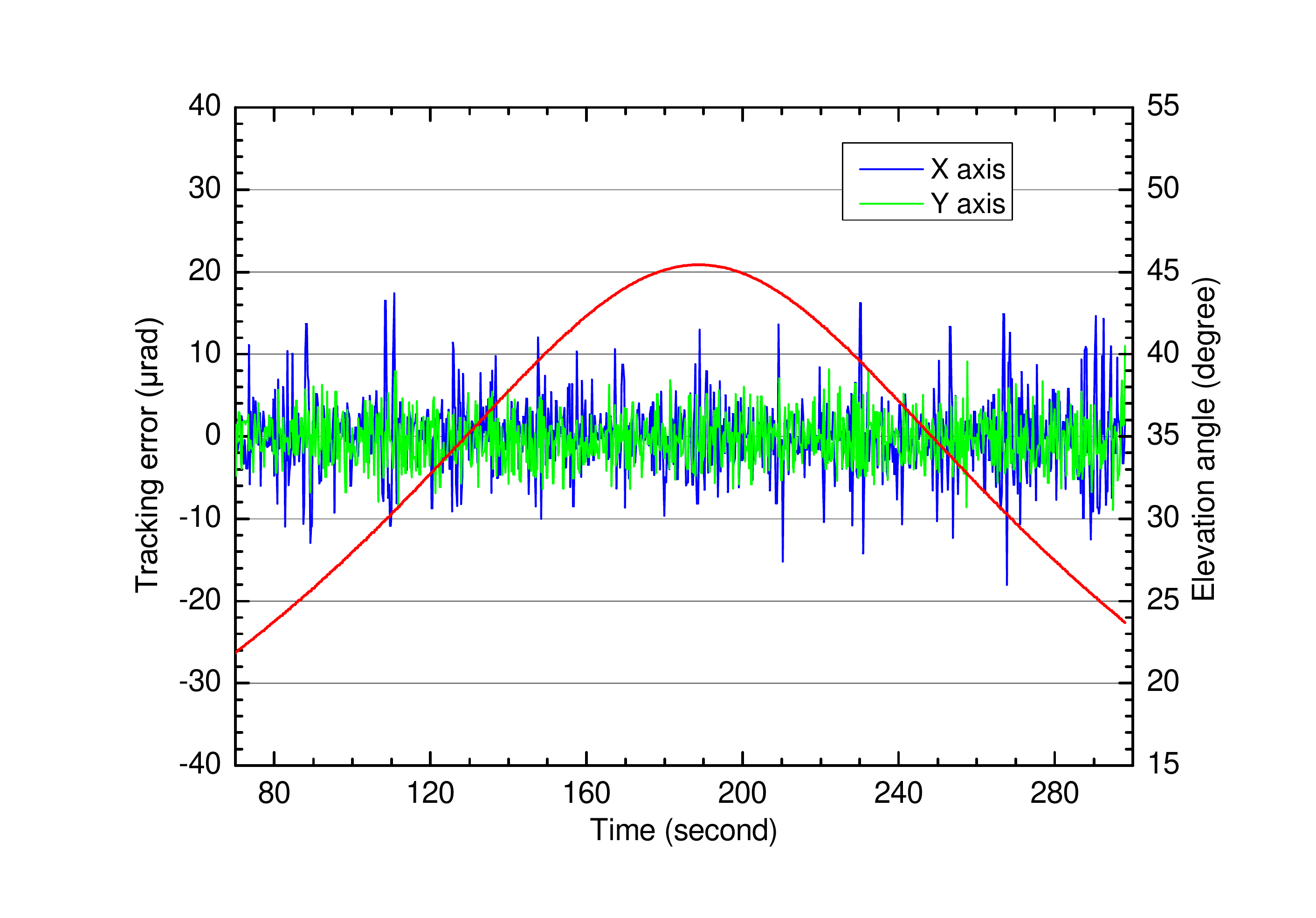}
\caption{Tracking errors (blue and green lines for the X- and Y-axes, respectively) during a tracking measurement at Shanghai in July 2019. The RMS errors for the tracking offset angles in the X- and Y-directions were \SI{4.41}{\micro\radian} and \SI{2.99}{\micro\radian}, respectively. These small offsets are negligible compared with the much larger receiving field of view. The red line represents the elevation angle from the ground station to the satellite. These data were selected during the efficient time period when the satellite was higher than about \SI{20}{\degree}.}
\label{fig:track}
\end{figure}

Several minutes before the satellite passage, the operator must load the trajectory file for the satellite into the controlling software of the ground station. The beacon light from the ground station then points to the direction at the beginning of the satellite track and waits for the satellite. When the satellite appears at this position, it sees the beacon from ground station and emits a green beacon back to the ground after close-loop positioning is finished at the satellite side. The ground station then acquires and tracks this green light to establish a stable and efficient optical link. The whole process is completed automatically on both sides, and the operator is only required to monitor the process. The tracking performance in a measurement at Shanghai in July 2019 is shown in Fig.~\ref{fig:track}. The tracking error was only several micro-radians after the optical link was established and stable.

\begin{table}[tb]
\centering
\caption{QKD data from 17 different orbits at six separate ground stations.}
\setlength{\tabcolsep}{1.5mm}{
\begin{tabular}{ccccccc}   
	\hline
	Site & Date & $A$ &$T$\,(\si{s}) & $S$\,(\si{kb}) & $QBER$ & $K$\,(\si{bits}) \\
	\hline
	\multirow{6}{*}{Shanghai} & 10/03/2018 & \ang{72} & \num{78}& \num{345.9} & 1.36\% & \num{63791} \\
	 & 10/23/2018 & \ang{78} & \num{106}& \num{301.1}  & 2.90\% & \num{25163} \\
	 & 04/26/2019 & \ang{50} & \num{114}& \num{172.5}  & 2.00\% & \num{15488} \\
	 & 05/01/2019 & \ang{84} & \num{107}& \num{213.0}  & 1.90\% & \num{31296} \\
	 & 05/04/2019 & \ang{38} & \num{168}& \num{296.9}  & 1.37\% & \num{51829} \\
	 & 05/22/2019 & \ang{80} & \num{71}& \num{225.2}  & 0.80\% & \num{45888} \\
	\hline
	\multirow{1}{*}{Beijing} & 08/07/2020 & \ang{52} & \num{110}& \num{657.7} & 1.20\% & \num{80640} \\
	\hline
	\multirow{3}{*}{Jinan} & 09/02/2020 & \ang{73} & \num{168}& \num{598.2} & 2.20\% & \num{70928} \\
	 & 09/06/2020 & \ang{81} & \num{227}& \num{661.5}  & 1.95\% & \num{124351} \\
	 & 09/11/2020 & \ang{41} & \num{245}& \num{624.9}  & 2.38\% & \num{68868} \\
	\hline
	\multirow{4}{*}{Weihai} & 05/17/2019 & \ang{65} & \num{174}& \num{312.5}  & 2.91\% & \num{33024} \\
	 & 05/21/2019 & \ang{56} & \num{280}& \num{490.4}  & 2.05\% & \num{67734} \\
	 & 05/23/2019 & \ang{33} & \num{259}& \num{316.6}  & 2.40\% & \num{12800} \\
	 & 07/01/2019 & \ang{30} & \num{345}& \num{526.6}  & 1.65\% & \num{93296} \\
	\hline
	\multirow{2}{*}{Mohe} 
	 & 08/24/2019 & \ang{55} & \num{175}& \num{262.2}  & 1.73\% & \num{45599} \\
	 & 08/25/2019 & \ang{85} & \num{186}& \num{330.8}  & 2.50\% & \num{31997} \\
	\hline
	\multirow{1}{*}{Lijiang} & 03/25/2019 & \ang{72} & \num{145}& \num{271.7}  & 2.11\% & \num{30735} \\
	\hline
\end{tabular}
}

\raggedright
$A$, highest altitude angle;$T$, efficient time; $S$, sifted key length for a single passage; $QBER$, quantum bit error rate for signal state; $K$, final key size applied by statistical fluctuation analysis \cite{zhang2017}.
\label{tab:QKDresult}
\end{table}

From 2018 to 2020, we have successfully carried out 17 tracks of QKD at six different ground stations, the details of which are summarized in Table~\ref{tab:QKDresult}.
From Oct 3, 2018 to May 22, 2019, we achieved six measurements at the ground station in Shanghai (31.1263°N, 121.5424°E, altitude 25 m). Due to the different weather conditions, the efficient time ranged from 71 to 168 seconds, and 1.81 Mb sifted keys were obtained with average QBER 0.8\% to 2.9\%.
In an efficient time of 110 seconds at the ground station in the center of Beijing (39.8853°N, 116.3514°E, altitude 120 m), 658 kb sifted keys were obtained, and the average QBER was 2\%.
At the ground station in Jinan (36.6768°N, 117.1233°E, altitude 85 m), three tracks were arranged for QKD. We obtained 1.88 Mb sifted keys, with the average QBER ranging from 1.95\% to 2.38\%, during an efficient time of 168 to 245 seconds. 
From May 17, 2019 to July 1, 2019, we achieved four measurements at the Weihai station (37.5340°N,122.0513°E, altitude 46 m), obtaining 1.65 Mb sifted key with an average QBER ranging from 0.8\% to 2.9\%, during an efficient time of 174 to 345 seconds.
In August 2019, in the northernmost village in Mohe county (53.4852°N, 122.3537°E, altitude 300 m), we successfully achieved two measurements and obtained 592 kb sifted keys with the average QBER ranging from 1.73\% to 2.50\%.
The last station was in Lijiang (26.6939°N, 100.0293°E, altitude 3233 m). In an efficient time of 145 seconds, we obtained 271 kb sifted keys with a QBER of 2.11\%.

For secure key generation, we used standard decoy-state QKD analysis method for data post-processing.
Considering the statistical fluctuations the failure probability is set to $10^{-5}$.
We obtained the following final key sizes: 239.7 kb for Shanghai, 80.6 kb for Beijing, 264.1 kb for Jinan, 206.9 kb for Weihai, 77.6 kb for Mohe, and 30.7 kb for Lijiang.
Actually, the portable quantum communication ground stations in Shanghai, Beijing, and Jinan have begun to provide unconditionally secure keys to users, including banks and a power-grid company, for data encryption in long-distance transmission.

\section{Conclusions and outlook}
In this paper, we have described a portable quantum communication ground station designed for the user's requirements of miniaturization and portability. Its weight has been reduced to less than 100 kg, and it can be installed and tested in 12 hours and can perform QKD with the satellite. This kind of ground station has significant advantages for the application scenarios that are difficult to cover with optical fibers, such as offshore islands and ships. We have successfully performed 17 QKD verification experiments in six different places, with the average final secure key being around 50 kb during one passage, based on these experiments. This kind of ground stations has more advantages for the application scenarios that are difficult to cover by optical fiber, such as offshore islands and ships. Such portable ground stations will greatly speed up the practical process of satellite-to-ground QKD and the construction of global quantum network \cite{Kimble_QInternet_Nature}. Our ground-station design is also valuable for the satellite optical communication terminals \cite{toyoshima:2008:laser_communication,kaushal:FSO:2016}.

\medskip
\noindent\textbf{Data availability.} 
The data that support the findings of this study are available from the corresponding authors on reasonable request.


\medskip
\noindent\textbf{Funding.} Strategic Priority Research Program on Space Science (Chinese Academy of Sciences),
National Key R\&D Program of China (2017YFA0303900 and 2018YFE0200600),
the Shanghai Municipal Science and Technology Major Project (2019SHZDZX01),
the Anhui Initiative in Quantum Information Technologies,
the National Natural Science Foundation of China (U1738203 and U1738204),
Key-Area Research and Development Program of Guangdong Province (2018B02038001),
and Anhui Province for Outstanding Youth (1808085J18).

\medskip
\noindent\textbf{Acknowledgments.} The authors thank our colleagues in Shanghai Institute of Technical Physics, National Space Science Center, and Xi’an Satellite Control Center for arrange the link opportunities of Micius satellite, and our friends in State Grid Information \& Telecommunication Co., Ltd, Jinan Institute of Quantum Technology,  Weihai Campus of Shandong University, Arctic Village in Mohe County of Heilongjiang Province and Gaomeigu Observational Station of Yunnan Observatory for implementation site support.

\medskip
\noindent\textbf{Competing interests.} The authors declare no conflicts of interest.

\medskip
\noindent\textbf{Contributions.} 
C.-Z.P. and J.-W.P. conceived the research. J.-G.R., M. A., H.-L.Y., C.-Z. P. and J.-W.P. designed the experiment. J.-G.R., H.-L.Y., J.Y., X.-J.L. Y.-Y.Y. and X.-J.Z. designed and developed the optical parts of the ground stations. J.-G.R., H.-L.Y., H.-J.X. and S.-Q.Z.designed and developed the mounts and tracking softwares, M.A., H.-L.Y., B.J., S.-K. L., W.-Q.C. and C.-Z.P. designed the electronics. J.-G.R., J.Y., S.-K.L.,W.-Q.C., W.-Y.L., C.-Z.P., J.-Y.W. and J.-W.P. designed and developed the payloads on the satellite. J.-G.R., H.-L.Y., J.Y., F.-Z.L., W.-Y.L., Y.-A.C., C.-Z.P. and J.-W.P. analysed the data and wrote the manuscript. All the authors contributed to the data collection, discussed the results and reviewed the manuscript. J.-W.P. supervised the whole project.

\end{document}